\documentclass[11pt]{article}
\usepackage{epsfig}
\usepackage{latexsym}
\usepackage{amssymb}
\begin{document}

\title{The Yang-Mills theory as a massless limit of a massive nonabelian
gauge model.}
\author{ A.A.Slavnov \thanks{E-mail:$~~$ slavnov@mi.ras.ru}
\\Steklov Mathematical Institute, Russian Academy
of Sciences\\ Gubkina st.8, GSP-1,119991, Moscow\\
                  and Moscow State University} \maketitle

\begin{abstract}
A gauge invariant infrared regularization of the Yang-Mills theory applicable
beyond perturbation theory is constructed.
\end{abstract}

\section {Introduction}

It is well known that the Yang-Mills theory cannot be obtained  as a
limit, when $m \rightarrow 0$, of the massive Yang-Mills model. The
longitudinal polarization of the massive vector field does not decouple
in the zero mass limit and gives a nonzero contribution \cite{SF}. It was
shown recently \cite{Sl1}, that a massive nonabelian gauge model, which
includes additional excitations, may be used as a gauge invariant
infrared regularization of the Yang-Mills theory. However the procedure
proposed in \cite{Sl1} may be used only in the framework of perturbation
theory with respect to the coupling constant. On the other hand in the
presence of infrared singularities a perturbation theory is not
applicable, although the regularization, proposed in \cite{Sl1} is still
useful for calculation of Green functions of gauge invariant operators.

In the present paper we propose a gauge invariant nonabelian model which
may be used as an infrared regularization of the Yang-Mills theory both in
the framework of the perturbation theory and beyond it.

\section{The massive nonabelian gauge invariant model.}

It was discussed in the papers (\cite{Sl2}, \cite{Sl3},\cite{QS}) that
impossibility to select a unique gauge beyond the perturbation theory is
not the intrinsic property of the Yang-Mills model, but is related to its
particular formulation.  Adding new excitations which decouple
asymptotically it is possible to quantize nonabelian gauge models in a
manifestly Lorentz invariant way both in perturbation theory and beyond
it.

Having that in mind we propose to use for the gauge invariant infrared
regularization of the Yang-Mills theory the following Lagrangian
\begin{eqnarray}
L=- \frac{1}{4}F_{\mu \nu}^aF_{\mu \nu}^a  -m^{-2}(D^2 \tilde{\phi})^*
(D^2\tilde{\phi}) +(D_{\mu}e)^* (D_{\mu}b)+(D_{\mu}b)^*(D_{\mu}e)\nonumber\\
+ \alpha^2(D_{\mu} \tilde{\phi})^*( D_{\mu} \tilde{\phi})-\alpha^2m^2(b^*e+e^*b)
\label{1}
\end{eqnarray}
where $\phi$ is a two component complex doublet, and
\begin{equation}
\tilde{\phi}= \phi- \hat{\mu}; \quad \hat{\mu}=(0, \mu \sqrt{2}g^{-1})
\label{2}
\end{equation}
$\mu$ is an arbitrary constant. $D_{\mu}$ denotes the usual covariant
derivative. To save the place we consider here the group $SU(2)$.

In the following we shall use the parametrization of $\phi$ in terms of Hermitean
components
\begin{equation}
\phi=(\frac{i \phi^1+ \phi^2}{\sqrt{2}}(1+ \frac{g}{2 \mu} \phi^0),
\frac{\phi^0- i \phi^3(1+g/(2 \mu) \phi^0)}{\sqrt{2}}) \label{3}
\end{equation}

The complex anticommuting scalar fields $b,e$ will be parameterized as
follows
\begin{eqnarray}
b=(\frac{ib^1+b^2}{\sqrt{2}},
\frac{b^0-ib^3}{\sqrt{2}})(1+ \frac{g}{2 \mu} \phi^0)\nonumber\\
e=(\frac{ie^1+e^2}{\sqrt{2}}, \frac{e^3}{\sqrt{2}})\label{4}
\end{eqnarray}
where the components $e^{\alpha}$ are Hermitean, and $b^{\alpha}$ are
antihermitean. This particular parametrization of the classical fields is
used as we want to get rid off the ambiguity in choosing the gauge for
quantization completely.

The Lagrangian (\ref{1}) is obviously invariant with respect to "shifted" gauge transformations
\begin{eqnarray}
A_{\mu}^a \rightarrow A_{\mu}^a+\partial_{\mu} \eta^a-g \epsilon^{abc}A_{\mu}^b \eta^c\nonumber\\
\phi^a \rightarrow \phi^a + \frac{g}{2} \epsilon^{abc}\phi^b \eta^c+ \eta^a \mu
+\frac{g^2}{4\mu}\phi^a \phi^b \eta^b\nonumber\\
\phi^0 \rightarrow \phi^0-\frac{g}{2}(\phi^a \eta^a)(1+\frac{g}{2\mu}\phi^0)\nonumber\\
b^a \rightarrow b^a+ \frac{g}{2} \epsilon^{abc}b^b \eta^c+
\frac{g}{2}b^0 \eta^a+\frac{g^2}{4 \mu}b^a \phi^b \eta^b\nonumber\\
e^a\rightarrow e^a+ \frac{g}{2}\epsilon^{abc}e^b \eta^c+ \frac{g}{2}e^0
\eta^a\nonumber\\
b^0 \rightarrow b^0-\frac{g}{2}b^a \eta^a+ \frac{g^2}{4\mu}(\phi^a \eta^a)b^0\nonumber\\
e^0 \rightarrow e^0-\frac{g}{2}e^a \eta^a.
 \label{5}
\end{eqnarray}

This Lagrangian is also invariant with respect to the supersymmetry transformations
\begin{eqnarray}
\phi \rightarrow \phi-b \epsilon\nonumber\\
e \rightarrow e- \frac{D^2(\phi- \hat{\mu})}{m^2} \epsilon\nonumber\\
b \rightarrow b \label{6}
\end{eqnarray}
where $\epsilon$ is a constant Hermitean anticommuting parameter. This
symmetry plays a crucial role in the proof of decoupling of unphysical
excitations. It holds for any $\alpha$, but for $\alpha=0$ these
transformations are also nilpotent. Note that for further discussion we
need only the existence of the conserved charge $Q$
 and nilpotency of the asymptotic charge $Q_0$, as the physical spectrum
 is determined by the asymptotic dynamics. (We use the standard scattering problem,
  assuming that the asymptotical spectrum coincides with the spectrum of the free
  Hamiltonian).
   In the case under
consideration the nilpotency of the asymptotic charge requires
 $\alpha=0$, and the massive theory with $\alpha \neq 0$ is gauge
 invariant but not unitary. It may seem strange as usually the gauge
 invariance is a sufficient condition of unitarity, because one
 can pass freely from a renormalizable gauge to the unitary one, where
 the spectrum includes only  physical excitations. In the present case
 there is no "unitary" gauge. Even in the gauge  $\phi^a=0$, there are
 unphysical excitations.

 For gauge transformations (\ref{5}) the gauge $\phi^a=0$ is admissible
both in perturbation theory and beyond it. Indeed, if $\phi^a=0$, then
under the gauge transformations (\ref{5}) the variables $\phi^a$ become
\begin{equation}
\delta \phi^a= \mu \eta^a
\label{7}
\end{equation}
and the condition $\phi^a=0$ implies that $\eta^a=0$. In the gauge
$\phi^a=0$ the variables $\phi^\alpha$ are parameterized as usual
\begin{equation}
\phi=(0, \frac{\phi^0}{\sqrt{2}})+\hat{\mu} \label{8}
\end{equation}
It is also obvious that for $\alpha \neq 0$ the Lagrangian (\ref{1})
describes  a massive vector field and does not produce infrared
singularities.

In terms of shifted variables the Lagrangian (\ref{1}) looks as follows
\begin{eqnarray}
L=- \frac{1}{4} F_{\mu \nu}^aF_{\mu \nu}^a-m^{-2}(D^2 \phi)^*(D^2 \phi)
+m^{-2}(D^2 \phi)^*(D^2 \hat{\mu})\nonumber\\
+m^{-2}(D^2 \hat{\mu})^*(D^2 \phi)-m^{-2}(D^2 \hat{\mu})^*(D^2 \hat{\mu})+(D_\mu e)^*(D_\mu b)\nonumber\\
+(D_\mu b)^*(D_\mu e)+ \alpha^2(D_\mu \phi)^*(D_\mu \phi)- \alpha^2(D_\mu \phi)^*(D_\mu \hat{\mu})\nonumber\\
- \alpha^2(D_\mu \hat{\mu})^*(D_\mu \phi)+ \alpha^2(D_\mu
\hat{\mu})^*(D_\mu \hat{\mu})- \alpha^2m^2(b^*e+e^*b) \label{10}
\end{eqnarray}
and the term
\begin{equation}
\alpha^2(D_\mu \hat{\mu})^*(D_\mu \hat{\mu})= \frac{\alpha^2 \mu^2}{2}A_
\mu^2 \label{10}
\end{equation}
produces the mass for the vector field.

The term
\begin{equation}
 m^{-2}(D^2 \hat{\mu})^*(D^2 \hat{\mu})=
\frac{\mu^2}{2m^2}[(\partial_\mu A_\mu)^2+\frac{g^2}{2}(A^2)^2]
\label{11a}
\end{equation}
makes the theory renormalizable for any $\alpha$. To avoid complications
due to the presence of the Yang-Mills dipole ghosts at $\alpha=0$ we put
$\mu^2=m^2$.
 The effective Lagrangian in the gauge
$\phi^a=0$ may be written as follows
\begin{equation}
L_{ef}=L+ \lambda^a \phi^a- \mu \bar{c}^a c^a \label{12}
\end{equation}
where $L$ is the Lagrangian (\ref{10}) and $\bar{c}^a, c^a$ are nondynamical ghost fields.

Invariance of the Lagrangian (\ref{10}) with respect to the gauge
transformation (\ref{5}) and the supersymmetry transformations (\ref{6})
makes the effective Lagrangian invariant with respect to the simultaneous
BRST transformations corresponding to (\ref{5}) and the supersymmetry
transformations (\ref{6}). If $s_1$ is the nilpotent operator,
corresponding to the simultaneous BRST and supersymmetry transformations
and the transformations of the fields $\lambda, \bar{c}, c$ are
\begin{equation}
s_1 \lambda^a=0; \quad s_1c^a=-\frac{g}{2}\epsilon_{abc}c^bc^c; \quad s_1
\bar{c}^a= \lambda^a \label{12a}
\end{equation}
 the effective Lagrangian may be also written in the form
\begin{equation}
L_{ef}=L+s_1 \bar{c}^a\phi^a=L(x)+ \lambda^a \phi^a- \bar{c}^a(\mu
c^a-b^a) \label{13}
\end{equation}

As it was indicated in the paper \cite{QS}, one can integrate over
$\bar{c},c$ in the path integral determining expectation value of any
operator corresponding to observable. It leads to the change $c^a=b^a
\mu^{-1}$. After such integration the effective Lagrangian becomes
invariant with respect to the transformations which are the sum of the
BRST transformations and the supersymmetry transformations (\ref{6}) with
$c^a=b^a \mu^{-1}$:
\begin{eqnarray}
 \delta A_\mu^a=D_\mu b^a \mu^{-1} \epsilon\nonumber\\
\delta \phi^a=0\nonumber\\
\delta \phi^0=-b^0(1+ \frac{g}{2\mu} \phi^0) \epsilon\nonumber\\
\delta e^a=(\frac{g}{2\mu} \epsilon^{abc}e^bb^c+
\frac{ge^0b^a}{2 \mu}+i\frac{D^2(\tilde{\phi})^a}{\mu^2})\epsilon\nonumber\\
\delta e^0=(-\frac{ge^ab^a}{2 \mu}-\frac{D^2(\tilde{\phi})^0}{\mu^2})\epsilon\nonumber\\
\delta b^a=\frac{g}{2\mu}\epsilon^{abc}b^bb^c\nonumber\\
\delta b^0=0 \label{14}
\end{eqnarray}
As the transformation (\ref{14}) preserves the condition $\phi^a=0$, we
omitted in this transformation all the terms proportional to $\phi^a$.
For the asymptotic Hamiltonian these transformations acquire the form
\begin{eqnarray}
\delta A_\mu^a=\partial_\mu b^a \mu^{-1} \epsilon\nonumber\\
\delta \phi^a=0\nonumber\\
\delta \phi^0=-b^0 \epsilon\nonumber\\
\delta e^a=\partial_\mu A_\mu^a \mu^{-1}\nonumber\\
\delta e^0=-\partial^2 \phi^0 \mu^{-2}\nonumber\\
\delta b^a=0\nonumber\\
\delta b^0=0. \label{13a}
\end{eqnarray}

 According to the Neuther theorem the
invariance with respect to the transformations (\ref{14}) generates a
conserved charge $Q$, and the physical asymptotic states may be chosen to
satisfy the equation
\begin{equation}
\hat{Q_0}| \psi>_{as}=0
\label{15}
\end{equation}
where
\begin{eqnarray}
Q_0= \int d^3x[(\partial_0 A_i^a- \partial_i A_0^a) \mu^{-1} \partial_i
b^a-\mu^{-1}\partial_\nu A_{\nu}^a \partial_0 b^a +\mu^{-2} \partial^2(
\partial_0 \phi^0) b^0- \mu^{-2} \partial_0 b^0 \partial^2( \phi^0)\nonumber\\- \mu
\alpha^2 b^aA_0^a] \label{16}
\end{eqnarray}
Due to the conservation of the Neuther charge this condition is invariant with respect to dynamics.

Using Ostrogradsky canonical variables for higher derivative systems one can rewrite the eq.
(\ref{16}) in the form
\begin{equation}
Q_0= \int d^3x[-( \partial_i p_i^ab^a \mu^{-1}+ \mu \alpha^2
b^aA_0^a)+\mu^{-1}p_0^a
\partial_0 b-(p_1- \alpha^2 \varphi_2)b+p_2 \partial_0b] \label{17}
\end{equation}
Here
\begin{eqnarray}
p_i^\alpha=\frac{\partial L}{\partial \dot{A}_i^\alpha}; \quad p_i=
\frac{\delta L}{\delta \dot{\phi}_i}-
\frac{d}{dt} \frac{\delta L}{\delta \ddot{\phi_i}};\nonumber\\
\quad H=p_i^\alpha \dot{A}_i^\alpha +p_b \dot{b}+p_e \dot{e}+p_1 \varphi_2+p_2 \partial_0 \varphi_2 -L\nonumber\\
p_i^a=\dot{A}_i^a-\partial_iA_0^a; \quad p_0^a=- \partial_\nu A_\nu^a;
\quad p_b^a=\dot{e}^a;
\quad p_e^a= \dot{b}^a\nonumber\\
\varphi_1= \phi^0; \quad \varphi_2= \dot{\phi}^0; \quad
p_2=-\mu^{-2}\partial^2 \phi^0;\nonumber\\ p_1= \mu^{-2} \partial^2
\dot{\phi}^0+\alpha^2 \dot{\phi}^0; \quad b^0=b, e^0=e, p_b=\dot{e},
p_e=\dot{b} \label{17a}
\end{eqnarray}
In these notations the free Hamiltonian looks as follows
\begin{eqnarray}
H_0= \frac{p_i^2}{2}- \frac{p_0^2}{2}-\partial_ip_i^aA_0^a+p_0^a
\partial_iA_i^a++1/4(F_{ij}^a)^2+p_b^ap_e^a+\partial_ib^a \partial_ie^a+\nonumber\\
 +p_bp_e +\partial_ib \partial_ie+p_1 \varphi_2-\frac{\mu^2}{2}p_2^2+p_2\triangle{
 \varphi_1}\nonumber\\
 -\frac{\alpha^2 \mu^2}{2}A_0^2+\frac{\alpha^2\mu^2}{2}A_i^2+\alpha^2
\mu^2be- \frac{\alpha^2}{2} \varphi_2^2+\frac{\alpha^2}{2} \partial_i
\varphi_1
\partial_i \phi_1 \label{18}
\end{eqnarray}

We want to prove that the Lagrangian(\ref{13}) really describes the
infrared regularization of the Yang-Mills theory. That means for $\alpha
\neq 0$ it corresponds to a massive gauge invariant theory and in the
limit $\alpha=0$ it describes the usual three dimensionally transversal
excitations of the Yang-Mills field. Of course for $\alpha \neq 0$ the
spectrum includes also some unphysical excitations.

In the limit $\alpha=0$ only the first two lines  of the eq.(\ref{18})
survive. They contain the terms depending only on the fields $A_0, A_i$
and corresponding canonical momenta which coincide with the usual
Yang-Mills Hamiltonian in the diagonal Feynman gauge and the fields $
\phi_0, b_0, e_0$. The fields $b_a,e_a$ play the role of the
Faddeev-Popov ghosts. By the usual arguments the longitudinal and
temporal components of the Yang-Mills field as well as the fields
$b_a,e_a$ decouple, and the physical states may include only transversal
components of the Yang-Mills field and variables corresponding to the
fields $\phi^0, b^0, e^0$. Below we shall show that the fields $ \phi^0,
b^0, e^0$ also decouple.

It follows from the eqs. (\ref{17a}) that the asymptotical momenta $p_1, p_2$ satisfy the free field equations
\begin{equation}
\partial^2{p_{1,2}}=0; \label{19}
\end{equation}
Therefore these momenta allow the standard expansion
\begin{eqnarray}
p_{1,2}(x)=(2\pi)^{-\frac{3}{2}} \int d^3k \frac{i
\sqrt{\omega}}{2}(a^+_{p_{1,2}}(\textbf{k})\exp\{ikx\}-a^-_{p_{1,2}}(\textbf{k})
\exp\{-ikx\})\nonumber\\
k_0=\omega=\sqrt{\textbf{k}^2} \label{20}
\end{eqnarray}

A similar expansion may be written for the asymptotic fields $b,e,
\dot{b},\dot{e}$
\begin{eqnarray}
b(e)(x)= (2 \pi)^{-\frac{3}{2}} \int d^3k \frac{1}{ \sqrt{2 \omega}}(b(e)^+(\textbf{k})
 \exp\{ikx\}+b(e)^-(\textbf{k}) \exp\{-ikx\})\nonumber\\
\dot{b}(\dot{e})(x)= (2 \pi)^{-\frac{3}{2}} \int d^3k \frac{i
\sqrt{\omega}}{ \sqrt{2}}(b(e)^+(\textbf{k}) \exp\{ikx\}
-b(e)^-(\textbf{k}) \exp\{-ikx\})\nonumber\\
k_0=\omega= \sqrt{\textbf{k}^2} \label{21}
\end{eqnarray}
The canonical variables describing the fields $\phi_0$ in general are not
oscillatory and cannot be presented by the operators in the usual Fock
space.

The asymptotic fields $\varphi_{1,2}, p_{1,2}$ satisfy the equations of
motion which follow from the Hamiltonian(\ref{18}):
\begin{equation}
\dot{\varphi_1}=\varphi_2; \quad
\dot{\varphi_2}+\mu^2p_2-\triangle{\varphi_1}=0; \quad -\dot{p_{2}}=p_1;
\quad -\dot{p_1}= \triangle{p_2} \label{22}
\end{equation}
In accordance with these equations the part of the asymptotic BRST charge
depending on variables $p_{1,2}$
\begin{equation}
\tilde{Q}_0= \int d^3x(p_2 \dot{b}+p_1b) \label{23}
\end{equation}
is conserved and does not depend on time. Therefore one can put $x_0$ in
the variables $p_{1,2}, b, \dot{b}$ equal to zero.

Using the equations (\ref{20}-\ref{22}) we get
\begin{eqnarray}
p_1(\textbf{x},0)=(2\pi)^{-\frac{3}{2}} \int d^3k
\frac{i\sqrt{\omega}}{\sqrt{2}}(a^+_1(\textbf{k})\exp\{-i\textbf{kx}\}-a^-_1(\textbf{k})
\exp\{i\textbf{kx}\})\nonumber\\
p_2(\textbf{x},0)=-(2\pi)^{-\frac{3}{2}} \int d^3k
\frac{1}{2\sqrt{\omega}}(a^+_1(\textbf{k})\exp\{-i\textbf{kx}\}+a^-_1(\textbf{k})
\exp\{i\textbf{kx}\}) \label{23a}
\end{eqnarray}
\begin{eqnarray}
b(e)(\textbf{x},0)=(2\pi)^{-\frac{3}{2}} \int d^3k \frac{1}{\sqrt{2
\omega}}(b(e)^+(\textbf{k},0) \exp \{-i\textbf{kx}\}+b(e)^-(\textbf{k})
\exp\{i\textbf{kx}\})\nonumber\\
\dot{b(e)}(\textbf{x},0)=i(2\pi)^{-\frac{3}{2}}\int d^3k
\frac{\sqrt{\omega}} {\sqrt{2}}(b(e)^+(\textbf{k}) \exp
\{-i\textbf{kx}\}-b(e)^-(\textbf{k}) \exp\{i\textbf{kx}\}) \label{24}
\end{eqnarray}
 Let us introduce the following combinations of the operators $\hat{p}_{1,2},
\hat{\varphi}_{1,2}, \hat{b}, \hat{e}$, which satisfy the commutation
relations of creation and annihilation operators
\begin{eqnarray}
\frac{\hat{p}_2(\textbf{k})\omega(\textbf{k})+i
\hat{p}_1(\textbf{k})}{\sqrt{2\omega}}=-\hat{a}^+(\textbf{k}); \quad
\frac{\hat{\phi}_1(\textbf{k})\omega(\textbf{k})+i\hat{\phi}_2(\textbf{k})}
{\sqrt{2\omega}}=\hat{\tilde{a}}^-(\textbf{k})\nonumber\\
\frac{\hat{p}_2(\textbf{k})\omega(\textbf{k})-i
\hat{p}_1(\textbf{k})}{\sqrt{2\omega}}=-\hat{a}^-(\textbf{k}); \quad
\frac{\hat{\phi}_1(\textbf{k})\omega(\textbf{k})-i
\hat{\phi}_2(\textbf{k})}{\sqrt{2\omega}}=\hat{\tilde{a}}^+(\textbf{k})\nonumber\\
\frac{\hat{b}(\textbf{k})\omega(\textbf{k})+i
\dot{\hat{b}}(\textbf{k})}{\sqrt{2\omega}}=-\hat{b}^+(\textbf{k}); \quad
\frac{e(\textbf{k}) \omega(\textbf{k})+
i\dot{\hat{e}}(\textbf{k})}{\sqrt{2\omega}}=\hat{e}^-(\textbf{k})\nonumber\\
\frac{\hat{b}(\textbf{k})\omega(\textbf{k})-i
\dot{\hat{b}}(\textbf{k})}{\sqrt{2\omega}}=-\hat{b}^-(\textbf{k}); \quad
\frac{\hat{e}(\textbf{k})\omega(\textbf{k})-i\dot{\hat{e}}(\textbf{k})}{\sqrt{2\omega}}=\hat{e}^+(\textbf{k})
\label{24a}
\end{eqnarray}
Note that noncommuting pairs are formed by the operators $\hat{a},
\hat{\tilde{a}}$ and $\hat{b}, \hat{e}$.

For oscillatory variables $p_{1,2}, b, e$ this definition coincides with
the standard one.

In terms of these operators the asymptotic BRST charge may be written as
follows
\begin{equation}
Q_0=i \int d^3k( \hat{a}^+(\textbf{k})\hat{b}^-(\textbf{k})-
\hat{b}^+(\textbf{k})\hat{a}^-(\textbf{k})) \label{27}
\end{equation}
I wish to emphasize that the states generated by the operators
$\hat{\tilde{a}}^+$ do not belong to the Fock space. However this fact is
irrelevant for the physical interpretation as the part of the Hamiltonian
(\ref{18}) $\hat{\tilde{H}}$, which depends on the variables
$b,e,\varphi_1, \varphi_2$ and conjugated momenta, is BRST exact: it may
be presented as the anticommutator of the BRST charge (\ref{23}) with
some operator $A$
\begin{equation}
\hat{\tilde{H}}_0=[\hat{Q}_0,\hat{A}]_+; \quad \hat{A}= \int d^3x(
\hat{\varphi}_2 \hat{\dot{e}}+\frac{\mu^2}{2} \hat{p}_2 \hat{e}+
\triangle{\hat{\varphi}}_1 \hat{e}) \label{28}
\end{equation}

It follows that the part of the Hamiltonian $\hat{\tilde{H}}$, which
depends on $\varphi_{1,2}, p_{1,2}, b, e$ does not contribute to the
expectation value calculated with the help of the physical states,
annihilated by $Q_0$. (For similar construction see \cite{HT}).
Therefore it is irrelevant for the energy of any physical state, and
we can define the vacuum as the vector annihilated by the operators
$\hat{a}^-, \hat{\tilde{a}}^-, \hat{b}^-, \hat{e}^-$.

We may introduce the operator $\hat{K}$ by the formula
\begin{equation}
\hat{K}= \int d^3x(\hat{\varphi}_1(\textbf{x},0)
\hat{\dot{e}}(\textbf{x},0)-\hat{\varphi}_2(\textbf{x},0)
\hat{e}(\textbf{x},0))=-i\int d^3k (\hat{\tilde{a}}^+(\textbf{k})
\hat{e}^-(\textbf{k})-\hat{e}^+(\textbf{k})\hat{\tilde{a}}^-(\textbf{k}))
\label{29}
\end{equation}
The anticommutator of $\hat{Q}_0$ and $ \hat{K}$ is proportional to
the number of unphysical modes generated by the operators
$\hat{a}^+, \hat{\tilde{a}}^+, \hat{b}^+, \hat{e}^+$. By the usual
arguments any vector satisfying the eq.(\ref{15}) can be presented
in the form
\begin{equation}
|\psi>_{phys}=|\psi>_{tr}+Q_0|\chi> \label{30}
\end{equation}
where $|\psi>_{tr}$ contains only excitations corresponding to the
transversal modes of the Yang-Mills field, and expectation value of any
observable, calculated with the help of the vectors $|\psi>_{phys}$
coincides with the expectation value calculated with the help of
$|\psi>_{tr}$.

It completes the proof of the fact that for $\alpha = 0$ the Lagrangian
(\ref{13}) produces the same expectation values for all observables
calculated with the help of of the physical vectors , satisfying the
eq.(\ref{15}) as the standard Yang-Mills theory. At the same time for
$\alpha \neq 0$ it describes a massive gauge invariant theory, which does
not have infrared singularities.

\section{Discussion}

In this paper we showed that the Lagrangian (\ref{10})may be uniquely
quantized irrespectively of using the perturbation theory with respect to
the coupling constant. For any value of the coupling constant it allows a
unique canonical quantization in the gauge $\phi^a=0$.

For $\alpha \neq 0$ it does not produce infrared singularities and is
gauge invariant. Introducing some ultraviolet gauge invariant
regularization, for example dimensional or higher covariant derivatives,
one get the gauge invariant regularization which makes all the quantities
finite.

For $\alpha=0$ it gives for the expectation values of observable
operators, calculated with the help of the physical vectors, annihilated
by the BRST charge the same result as the usual Yang-Mills Lagrangian. Of
course, if the calculations are performed in the framework of
perturbation theory, for $\alpha=0$ the infrared singularities reappear.
However this Lagrangian may serve as a starting point for nonperturbative
calculations, in particular for calculations explaining the phenomenon of
quark confinement.

\textbf{Acknoledgements}

I wish to thank Andrea Quadri for reading the manuscript and helpful
comments. This paper was supported in part by RFBR under grants
11-01-00296a and 11-01-12037 ofi-m-2011, by the grant of support of
leading scientific schools NS-4612.2012.1 and by the program "Nonlinear
dynamics".

\end{document}